\documentclass[pre,aps,amsmath,showpacs]{revtex4}
\begin{document}
\title{Conventional methods fail to measure $c_p(\omega)$ of glass-forming liquids}
\author{Tage Christensen, Niels Boye Olsen, and Jeppe C. Dyre}
\affiliation{DNRF centre  ``Glass and Time,'' IMFUFA, Department of Sciences, Roskilde University, Postbox 260, DK-4000 Roskilde, Denmark}
\date{\today}
\newcommand{\iomt}{i\omega\tau}\newcommand{\bu}{{\bf u}}\newcommand{\br}{{\bf r}}\newcommand{\bnul}{{\bf 0}}\newcommand{\bnab}{{\bf\nabla}}\newcommand{\pa}{\partial}\newcommand{\rn}{\rho_0}\newcommand{\dt}{\delta T}\newcommand{\bv}{\beta_V}\newcommand{\bj}{{\bf j}_Q}\newcommand{\lap}{\nabla^2}\newcommand{\divu}{\bnab\cdot\bu}\newcommand{\sk}{\sinh (k}\newcommand{\ck}{\cosh (k}\newcommand{\mma}{\bf M\it}\newcommand{\tma}{\bf T\it}\newcommand{\pma}{\bf P}\newcommand{\tx}{\tilde x}\newcommand{\ty}{\tilde y}\newcommand{\tL}{\tilde L}\newcommand{\tJ}{\tilde J}\newcommand{\tit}{\tilde t}\newcommand{\tf}{\tilde f}\newcommand{\tu}{\tilde u}\newcommand{\tal}{\tilde\alpha}\newcommand{\tsig}{\tilde\sigma_{xx}}\newcommand{\dtq}{\delta\tilde q}\newcommand{\ttemp}{\tilde T}\newcommand{\tg}{\tilde g}\newcommand{\tc}{\tilde c}\newcommand{\dtt}{\delta\tilde T}\newcommand{\dv}{\Delta V}

\begin{abstract}The specific heat is frequency dependent in highly viscous liquids. By solving the full one-dimensional thermo-viscoelastic problem analytically it is shown that, because of thermal expansion and the fact that mechanical stresses relax on the same time scale as the enthalpy relaxes, the plane thermal-wave method of does not measure the isobaric frequency-dependent specific heat $c_p(\omega)$.  This method rather measures a ``longitudinal'' frequency-dependent specific heat, a quantity defined and detailed here that is in-between $c_p(\omega)$ and $c_v(\omega)$ . This result means that no wide-frequency measurements of $c_p(\omega)$ on liquids approaching the calorimetric glass transition exist. We briefly discuss consequences for experiment.\end{abstract}


\maketitle

\section{Introduction}

All liquids are viscoelastic, i.e., flow only on long time scales, whereas they are elastic on short time scales. This fact was first described quantitatively in 1867 by James Clerk Maxwell \cite{max67}. If $\eta$ is the (dc) shear viscosity and $G_\infty$ the instantaneous shear modulus, the Maxwell model implies that the characteristic time $\tau$ separating ``long'' and ``short'' times is given by 

\begin{equation}\label{1a}
\tau\,=\,\frac{\eta}{G_\infty}\,.
\end{equation}
For liquids like ambient water $\tau$ is in the picosecond range; for highly viscous liquids $\tau$ can be seconds, hours, or even months.

When a liquid is supercooled, its viscosity increases dramatically upon continued cooling; since $G_\infty$ is not nearly as temperature dependent, the relaxation time $\tau$ also increases enormously. At some point -- if crystallization is avoided -- the liquid freezes into a glass. The glass transition takes place when the liquid is unable to fully equilibrate on the experimental time scale \cite{kau48}. The properties of viscous liquids approaching the glass transition are far from well understood \cite{har76,bra85,ang91,ang00,das04,dyr06}. A viscous liquid is characterized by its relaxation time which is the time it takes for the system to relax to equilibrium after an externally imposed disturbance. The disturbance may be electrical, thermal, or mechanical. The relaxation time depends somewhat on which quantity is measured, but as a first approximation all relaxation times may be regarded as identical. Since shear stresses also relax on this time scale, the liquid relaxation time is basically the Maxwell $\tau$ of Eq. (\ref{1a}). Indeed, numerous experiments on viscous liquids confirm that the Maxwell relaxation time determines the rate of molecular motion beyond pure vibration \cite{har76,bra85,ang91,ang00}. 

When probed by a linear-response experiment on time scales shorter than $\tau$, viscous liquids exhibit a solid-like response. As an example, consider dielectric relaxation. This is an experimental technique where accurate data nowadays are easily obtained over 9 or more decades \cite{lun00,kre02}. Upon increase of the angular frequency $\omega$, the dielectric constant $\epsilon(\omega)$ drops from its value when $\omega\tau\ll 1$ to a lower value at high frequencies ($\omega\tau\gg 1$) where dipolar rotations no longer contribute. The high-frequency dielectric constant corresponds to that observed in the glass where the molecules are basically frozen. Another example is the compressibility. This quantity is also frequency dependent and is also larger at low frequencies ($\omega\tau\ll 1$) than at high ($\omega\tau\gg 1$) \cite{chr94}. Again, the low compressibility observed at high frequencies reflects glassy behavior.

It has been known for almost a century that the specific heat is higher in the liquid phase than in the glass -- indeed, the glass transition is often identified from specific heat measurements \cite{hod94}. It thus appears that the specific heat in the viscous liquid phase must be frequency dependent. This possibility was discussed briefly by Angell and Torell in 1983 \cite{ang83}, but was also implicit in earlier theoretical papers \cite{roe77,moy78,moy81}. The frequency-dependent specific heat of a viscous liquid was first measured in 1985 by Birge and Nagel and by Christensen in two independent works utilizing different methods \cite{bir85,bir86,chr85}. By coincidence both works measured on glycerol, but the results were not consistent (a discrepancy that is still unsettled, incidentally). We here focus on the Birge-Nagel method; it covers a much wider frequency range than that of Ref. \cite{chr85}. 

Perhaps because of its unusual thermodynamic reference there were early doubts about the nature of the frequency-dependent specific heat \cite{oxt86,zwa88,jac90}. These critiques were based on the assumption coming from generalized hydrodynamics that, although transport coefficients may well be frequency dependent, linear constitutive quantities do not depend on frequency. This is an assumption, however, and not based on first principles. Today few researchers doubt that the frequency-dependent specific heat is a standard linear-response quantity; in fact, it is straightforward to derive a fluctuation-dissipation theorem for this quantity \cite{nie96}. Thus just as for any other linear response, upon a periodic small time-dependent temperature variation $\delta T(t)$, the enthalpy variation may be written as a convolution integral over the past temperature history. The frequency-dependent specific heat is the Laplace transform of the kernel of this convolution integral and its imaginary part determines the dissipation (i.e., free energy loss) \cite{bir85,chr85,nie96}. That the specific heat is frequency dependent is obvious in simple energy-based master-equation models \cite{nie96,bis05}; in fact even the simple one-dimensional asymmetric double-well potential has a frequency-dependent specific heat. Recent works brought further insights in this continuously developing field, both as regards novel experimental techniques \cite{pao00,hut00,jeo01,car02,hon02,mae02} and theory including computer simulations \cite{har01,sch01,yu04,cha05}.

The method of Birge and Nagel \cite{bir86,bir87} is based on so-called thermal effusion. A thin metal film evaporated on a solid glass substrate in the form of a plane slab is immersed in the liquid. The metal film acts both as heat generator and thermometer \cite{bir86,cah90,men96,moo96,bir97,jon98}. An applied oscillatory electric current at angular frequency $\omega_e$ generates oscillating Joule heating at the frequency $\omega=2\omega_e$ due to the electric resistance of the metal film. The generated power ``effuses'' as thermal waves into the liquid and into the supporting slab. The resulting temperature oscillations depend on both specific heats and heat conductivities. The amplitude and phase of the temperature oscillations are detected by the clever so-called $3\omega_e$-technique \cite{bir87,jon98}: Since the electric resistance of the film is temperature dependent, the resistance is perturbed at frequency $2\omega_e$ in concert with temperature. The detected voltage across the film thus contains a $3\omega_e$ component proportional to temperature arising from the product of varying current and resistance. The temperature is measured at the heat-producing film surface itself. If the thermal wavelength is short compared to the lateral dimensions of the film, the film is effectively an isothermal surface in space (not in time, of course). The heat current is orthogonal to this surface, implying that the thermal waves are planar and that boundary effects can be neglected.  In the periodic situation all fields vary with time proportional to $\exp(st)$ where $s=\pm i\omega$ depending on convention \cite{note1}. By solving the heat diffusion equation Birge and Nagel arrrived at the following expression for the (area-specific) thermal impedance of the liquid, $Z$, i.e., the ratio between the complex amplitudes of temperature, $\delta T(\omega)$, and heat current density, $j(\omega)$: 

\begin{equation}\label{0a}
Z\,\equiv\,\frac{\delta T(\omega)}{j(\omega)}\,=\,\left(sc_p\lambda\right)^{-1/2}\,.
\end{equation}
Here $\lambda$ is the heat conductivity and $c_p$ is the (generally frequency-dependent) specific heat per unit volume. In this paper we prefer to discuss the reciprocal function, the thermal admittance $Y\equiv 1/Z$ which is additive when taking the thermal response of the substrate into account: $Y=\left(sc_1\lambda_1\right)^{1/2}+\left(sc_2\lambda_2\right)^{1/2}$ where index $1$ refers to liquid properties and index $2$ to those of the supporting slab. Denoting the complex amplitude of the total power by $P(\omega)$, the total admittance $Y_{tot}=P(\omega)/\delta T(\omega)$ scales with the area $A$ of the film, $Y_{tot}=YA$. 

The terms ``thermal impedance'' and ``thermal admittance'' come, of course, from the close analogy between flowing electricity and flowing heat \cite{car59}. Temperature is analogous to potential and heat current to electric current. Thus Fourier's law becomes equivalent to Ohm's law, and the definition of heat capacity (specific heat) becomes equivalent to the definition of electric capacity. Even the dissipation of the free energy -- being proportional to the real part of the impedance -- follows from this scheme, justifying the terminology. 

According to Eq. (\ref{0a}) the bulk property measured by the Birge-Nagel experiment is the effusivity $\sqrt{\lambda c}$ \cite{min03,ben04,kra02}. If the heat conductivity is frequency independent, as was generally assumed and recently confirmed \cite{min01,ben03}, the method yields information about the frequency-dependent specific heat. The main conclusion of the present paper, however, is that the experiment {\it does not measure the isobaric specific heat}. This is because the ordinary heat diffusion equation fails for highly viscous liquids. The precise nature of the specific heat measured can only be clarified from a detailed analysis. This analysis is the subject of the paper. A brief description of the problem is the following: The ordinary (dc) specific heat exists in isochoric, $c_V$, and isobaric, $c_p$, versions. Similarly, there are two ac specific heats, $c_V(\omega)$ and $c_p(\omega)$. Since the pressure is usually ambient pressure, one would {\it a priori} expect that experiments measure $c_p(\omega)$ which, indeed, is what is always reported in the literature. There is a problem, however, which implies that $c_p(\omega)$ is not the quantity measured by the plane-plate experiment. The problem is that the coupling between the temperature field and the displacement field induced by thermal expansion implies that the ordinary heat diffusion equation fails. In other situations this can be ignored, but a subtle combination of two physical conditions prevailing at the glass transition renders the usual approach invalid: First, the quantity $(c_p-c_V)/c_V$ has a non-negligible value in the viscous liquid state. Secondly, upon increasing frequency the ratio of shear to bulk modulus increases to a non-negligible value. Physically what happens is that the high viscosity hinders the release of stresses induced by the thermal wave. Thermal stresses relax on the very time scale that experiments focus on. This implies that the specific heat is not measured at isobaric conditions; the stress tensor is not proportional to the unit tensor because the liquid is not in hydrostatic equilibrium. 

For solids the difference between $c_p$ and $c_V$ is small and can usually be neglected. For less-viscous liquids internal stresses are quickly released and the stress tensor is proportional to the unit tensor to a good approximation. The problem of measuring $c_p(\omega)$ reliably arises only close to the glass transition. Once the problem is recognized, the following questions arise: Which quantity is measured, and how should one proceed to obtain reliable $c_p(\omega)$ data? In this paper we present the formalism needed to answer these questions by treating the one-dimensional case in detail. From first principles we derive the results for the adiabatic boundary condition that were simply stated in Ref. \cite{chr97}; these results are supplemented by results for the isothermal boundary condition. After reviewing the basic equations governing thermo-viscoelasticity we derive the general solution of the one-dimensional case relevant for understanding the results of the Birge-Nagel plane-plate method. The solution is based on the transfer matrix formalism, a little-known but very convenient formalism which is introduced by first discussing the simpler special case of no thermo-mechanical coupling. Based on the general solution the thermal admittance is worked out for several cases corresponding to various experimentally relevant boundary conditions (adiabatic or isothermal, mechanically clamped or free). Finally, the results are discussed in light of what they tell us about the prospects of obtaining reliable $c_p(\omega)$ data by means of the plane plate method or otherwise.

\section{Basic equations of thermo-viscoelasticity}
\subsection{Thermoelasticity}

Thermoviscoelasticity describes the coupling between thermal and mechanical deviations from equilibrium. We shall only discuss the linear case which is well understood on the phenomenological level. To illustrate the general method we first discuss the simpler case where all linear-response coefficients are frequency independent. This is referred to as thermoelasticity, a theory that describes ordinary solids well \cite{chr82}.

The linearity requirement implies that the system is assumed to be infinitesimally close to equilibrium. The mass density is denoted by $\rho$, its average by $\rn$; the temperature is $T$ with the average temperature $T_0$. Deviations from equilibrium are quantified in terms of the displacement field $\bu=\bu(\br,t)$, temperature variation by $\dt(\br,t)=T(\br,t)-T_0$. The equations governing these variables involve the isothermal bulk modulus $K_T$ and the shear modulus $G$ (isothermal and adiabatic shear moduli are identical). Recall that these quantities are defined as follows: If $V$ is volume and $p$ pressure, the isothermal bulk modulus (inverse compressibility) is defined by

\begin{equation}\label{1}
K_T\,\equiv\,-V\left(\frac{\pa p}{\pa V}\right)_T\,.
\end{equation}
The shear modulus $G$ is by definition the proportionality between an off-diagonal component of the stress tensor and the corresponding shear deformation. 

If the heat current density is denoted by $\bj$, the basic thermoelastic equations involve the following three constitutive quantities: The heat conductivity $\lambda$ defined via $\bj=-\lambda\bnab\dt$, the isochoric specific heat per unit volume $c_V$, and the isochoric ``pressure-temperature coefficient'' $\bv$ defined by

\begin{equation}\label{2}
\bv\,\equiv\,\left(\frac{\pa p}{\pa T}\right)_V\,.
\end{equation}
The thermoelastic ``equations of motion'' are \cite{chr82,lan86,now86,chr97}

\begin{eqnarray}
\rn\frac{\pa^2\bu}{\pa t^2}\,&=&\,\left(K_T+\frac{4}{3}G\right)\bnab(\divu)-G\bnab\times(\bnab\times\bu)-\bv\bnab\dt\label{3}\\c_V\frac{\pa\,\dt}{\pa t}&+&\bv T_0\frac{\pa}{\pa t}\left(\divu\right)\,=\,\lambda\lap\dt\label{4}\,.
\end{eqnarray}
For completeness, we briefly sketch how these equations are derived: If $\pa_i$ is the derivative with respect to the i'th spatial coordinate $x_i$ and $u_i$ is the i'th component of the displacement vector $\bu$, the stain tensor $\epsilon_{ij}$ is defined by $\epsilon_{ij}=(\pa_iu_j+\pa_ju_i)/2$. Using the Einstein summation convention and denoting the stress tensor by $\sigma_{ij}$, we first note that the relevant constitutive relation is the Duhamel-Neumann relation \cite{lan86,now86},

\begin{equation}\label{5}
\sigma_{ij}\,=\,K_T\epsilon_{kk}\delta_{ij}+2G\left(\epsilon_{ij}-\frac{1}{3}\epsilon_{kk}\delta_{ij}\right)-\bv\dt\delta_{ij}\,.
\end{equation}
Considering a few special cases quickly convinces one that this equation is correct: If temperature is constant and the deformation is isotropic, only the first term contributes; thus Eq. (\ref{5}) reproduces the definition of the isothermal bulk modulus because $\epsilon_{kk}=\divu$ is the relative volume change and the pressure is the negative diagonal element of the stress tensor. If temperature is constant and a pure shear deformation is considered, the second term gives the definition of the shear modulus. Finally, if there are no mechanical deformations, the last term reduces to the definition of $\bv$, Eq. (\ref{2}). Now, Newton's second law for a small volume element $\dv$ is: $\rho\dv \ddot{u}_i=F_i$. Here $F_i$ is the i'th component of the force which is $\dv$ times the divergence of the stress tensor: $F_i=\dv\partial_j\sigma_{ij}$. Equation (\ref{3}) follows by substituting Eq. (\ref{5}) into Newton's second law, because $\rho$ to lowest order may be replaced by $\rn$. Equation (\ref{4}) is based on the following thermodynamic relation for the volume element $\dv$: $\delta S=(\pa S/\pa T)_V\dt+(\pa S/\pa V)_T\delta(\dv)=(\dv c_V/T_0)\dt+\bv\delta(\dv)$ where the last equality follows from the definition of $c_V$ and the Maxwell relation $(\pa S/\pa V)_T=\bv$. For the entropy per unit volume $s\equiv S/\dv$, since the relative volume change is the divergence of the displacement field, $\delta(\dv)/\dv=\bnab\cdot\bu$, we get $\delta s=(c_V/T_0)\dt+\bv \bnab\cdot\bu$. Entropy is conserved to first order: $(\pa s/\pa t)+\bnab\cdot{\bf j}_s=0$ where ${\bf j}_s=-(\lambda/T_0)\bnab(\delta T)$ is the entropy current density. Thus to lowest order we have $(\pa s/\pa t)=(c_V/T_0)(\pa \delta T/\pa t)+\bv\pa(\bnab\cdot\bu)/\pa t=\lambda/T_0\nabla^2(\dt)$ which is the required Eq. (\ref{4}).

Before proceeding we note that if $\alpha_p$ is the isobaric thermal expansion coefficient ($\alpha_p\equiv(1/V)(\pa V/\pa T)_p$), the mathematical identity $(\pa p/\pa T)_V(\pa T/\pa V)_p(\pa V/\pa p)_T=-1$ implies that

\begin{equation}\label{6}
\bv\,=\,\alpha_p\, K_T\,.
\end{equation}
In particular, if there is no thermal expansion upon heating isobarically ($\alpha_p=0$), one has $\bv=0$. In this case Eqs. (\ref{3}) and (\ref{4}) decouple and reduce to the ordinary elastic equation of motion and Fourier's equation for heat conduction, respectively. Thus the coupling between mechanics and thermodynamics arises only when there is nonzero thermal expansion -- which is indeed intuitively obvious.

\subsection{General thermoviscoelastic equations}

We now turn to the general linear thermoviscoelastic case. According to the ``correspondence principle'' \cite{chr82}, linear time-dependent equations are arrived at from the thermoelastic equations (\ref{3}) and (\ref{4}) by replacing each time derivative by an $s$ factor:

\begin{eqnarray}
&\rn&s^2\,  \bu\,=\,\left(K_T+\frac{4}{3}G\right)\bnab(\divu)-G\bnab\times(\bnab\times\bu)-\bv\bnab\dt\label{6b}\\&c_V&s\,\dt\,+\,\bv T_0s\,\divu\,=\,\lambda\lap\dt\label{6c}\,.
\end{eqnarray}
In these equations the constitutive coefficients $K_T$, $G$, $\bv$, $c_V$ and $\lambda$ are generally frequency dependent, but for simplicity we shall not explicitly indicate this. In this way the theory is generalized from thermoelasticity to thermoviscoelasticity. It is important to note that the symmetry of the constitutive equations are maintained; thus the Maxwell relations become Onsager relations \cite{mei59,ber78,moy78}. From here on we exclusively use the description in the frequency domain.

\section{Full analytic solution of the one-dimensional case} 

This section presents the solution of the one-dimensional case, i.e., the case where all functions vary only in one spatial direction $x$. The transfer matrix formulation of the solution given below is convenient in practice for discussing various experimentally relevant boundary conditions \cite{car59}. 

\subsection{Case of no thermo-mechanical coupling}

As an introduction to the transfer matrix technique we first solve the case of no thermo-mechanical coupling ($\bv=0$) which, although much simpler than the general case, maintains an independent interest. If  $c\equiv c_V$ (=$c_p$ when $\bv=0$) is the specific heat per unit volume, $\lambda$ the heat conductivity, and $j$ the $x$-component of the heat current $\bj$, the equations for temperature and heat current are

\begin{eqnarray}\label{7}
c\,\frac{\pa\dt}{\pa t}\,&=&\,-\frac{\pa j}{\pa x}\nonumber\\j\,&=&\,-\lambda\,\frac{\pa\dt}{\pa x}\,.
\end{eqnarray}
In a steady-state periodic situation where all variables vary in time propotionally to $\exp(st)$ these equations become

\begin{eqnarray}\label{8}c\,s\, \dt\,&=&\,-\frac{\pa j}{\pa x}\nonumber\\j\,&=&\,-\lambda\,\frac{\pa\dt}{\pa x}\,.
\end{eqnarray}
It is convenient to discuss the problem in terms of the two variables $\dt$ and $j$, although one usually focuses on the heat-diffusion equation which results from combining the two equations: $cs\dt=\lambda\pa_x^2\dt$. The general solution of this equation is $\dt(x)=c_1\exp(kx)+c_2\exp(-kx)$ where $c_1$ and $c_2$ are integration constants and

\begin{equation}\label{9}
k^2\,=\,\frac{c\,s}{\lambda}\,.
\end{equation}
Now $j(x)$ is found from $\dt(x)$: $j(x)=-\lambda k[ c_1\exp(kx)-c_2\exp(- kx)]$. The results may be summarized as follows:

\begin{equation}\label{10}
\begin{pmatrix}\dt(x) \\ J(x)\end{pmatrix}\,=\,\mma(x)\,\begin{pmatrix}c_1 \\ c_2\end{pmatrix}\,,
\end{equation}
where

\begin{equation}\label{11}
\mma(x)\,=\,\begin{pmatrix}e^{kx} & e^{-kx}\\-\lambda k e^{kx} & \lambda k e^{-kx}\end{pmatrix}\,.
\end{equation}
This formulation allows one to establish the connection between temperature- and heat-current amplitudes at one point, say $x=0$, and at the point at $x$: By eliminating the integration constants one arrives at

\begin{equation}\label{12}
\begin{pmatrix}\dt(x) \\ j(x)\end{pmatrix}\,=\,\tma(x)\,\begin{pmatrix}\dt(0) \\ j(0)\end{pmatrix}\,,
\end{equation}
where the ``transfer matrix'' $\tma(x)$ is defined by 

\begin{equation}\label{13}
\tma(x)\,=\,\mma(x)\mma({\rm 0})^{\rm -1}\,.
\end{equation}
By straightforward calculation one finds that the transfer matrix is given \cite{car59} by

\begin{equation}\label{14}
\tma(x)\,=\,\begin{pmatrix}\ck x) & -\sk x)/(\lambda k)\\-\lambda k\sk x) & \ck x)\end{pmatrix}\,.
\end{equation}
Further calculation shows that the consistency requirement $\tma(x)\tma(y)=\tma(x+y)$ is fulfilled; in particular $\tma({\rm 0})={\rm 1}$ and $\tma(-x)=\tma(x)^{\rm -1}$.

From the general solution various special cases are easily derived. Suppose the sample is located between $x=0$ and $x=L$ and heated periodically at $x=0$. To calculate the frequency-dependent (area-specific) ``thermal admittance'' $Y$ experienced from the heating side of the sample, $Y\equiv j(0)/\dt(0)$, we note that there are two experimentally relevant boundary conditions referring to the sample end at $x=L$: the adiabatic boundary condition, $j(L)=0$, and the isothermal boundary condition, $\dt(L)=0$. In the adiabatic case Eqs. (\ref{12}) and (\ref{14}) imply that $0=j(L)=-\lambda k\sk L)\dt(0)+ \ck L)j(0)$ which implies

\begin{equation}\label{15}
Y_{\rm adiab}\,=\,\lambda k\tanh(k L)\,.
\end{equation}
On the other hand, the isothermal boundary condition at $x=L$ is $0=\dt(L)=\ck L)\dt(0)-\sk L)J(0)/(\lambda k)$ which implies

\begin{equation}\label{16}
Y_{\rm isoth}\,=\,\lambda k\coth(k L)\,.
\end{equation}

Before proceeding to the general thermoviscoelastic case, we identify the limits of samples that are thick and thin, respectively, compared to the frequency-dependent thermal diffusion length: The complex $k$ defines a (complex) characteristic length via $l_D\equiv 1/k$. If $L\gg |l_D|$ one is in the ``thermally thick limit'' \cite{cae98}, a case that applies at high frequencies where the thermal wave does not reach the end of the sample. In this case the two solutions become identical:

\begin{equation}\label{17}Y_{\rm adiab}\,=\,Y_{\rm isoth}\,=\,\lambda k\,=\,(s\lambda c)^{1/2}\,\,\,(|kL|\gg 1)\,.
\end{equation}
In the ``thermally thin limit'' at low frequencies \cite{cae98}, $L\ll |l_D|$ , the adiabatic boundary condition leads to

\begin{equation}\label{18}
Y_{\rm adiab}\,=\,\lambda k^2\,L = scL\,\,\,(|kL|\ll 1)\,,
\end{equation}
corresponding to a capacitor. For the isothermal boundary condition the admittance becomes frequency independent:

\begin{equation}\label{19}
Y_{\rm isoth}\,=\,\frac{\lambda}{L}\,\,\,(|kL|\ll 1)\,.
\end{equation}
In this limit the magnitude of the heat current is solely determined by the temperature difference between the points $x=0$ and $x=L$, corresponding to the static limit of ordinary heat conduction.

\subsection{Full one-dimensional thermo-viscoelastic problem}

We now proceed along the above lines to determine the $4\times 4$ transfer matrix for the full one-dimensional thermo-viscoelastic problem, restricting ourselves to frequencies much lower than those where sound waves become important, the inertia-free limit. All relevant fields depend only on the $x$-coordinate and all displacement is in the $x$-direction. This does not mean that, e.g., $\sigma_{yy}$ is zero -- on the contrary, the stress tensor adjusts itself to be consistent with the clamping of all motion not in the exact $x$-direction. Physically, this situation corresponds to a box-shaped sample where the boundary conditions on the faces perpendicular to the heating plane are {\it adiabatic} and {\it sliding}. More realistically, the one-dimensional case to a good approximation describes the case where the dimensions of the plane heater are much larger than the relevant thermal wavelength.

We still assume periodically varying fields $\propto\exp(st)$. The first of the four coupled equations to be solved comes from Eq. (\ref{6b}). In the inertia-free limit the left-hand side is zero. If the $x$-coordinate of the displacement vector $\bu$ is denoted by $u$, since $\bnab\times\bu={\bf 0}$, this equation when integrated over $x$ implies (where $a_1$ is an integration constant)

\begin{equation}\label{20}
\pa_x u\,=\,\frac{\bv}{K_T+4G/3}\dt\,+\,a_1\,.
\end{equation}
Just as with all amplitudes and coefficients here and below, the integration constants generally depend on frequency. The next equation is arrived at by substituting Eq. (\ref{20}) into Eq. (\ref{6c}) which leads to 

\begin{equation}\label{21}
\lambda\,\pa_x^2\dt\,=\,s c_l\,\dt\,+\,s\bv T_0\, a_1\,,
\end{equation}
where the ``longitudinal heat capacity'' $c_l$ (a name justified by the results derived below) is defined by

\begin{equation}\label{22}
c_l\,=\,c_V\,+\,T_0\frac{\bv^2}{K_T+4G/3}\,.
\end{equation}
In passing we note that, utilizing Eq. (\ref{6}) and the well-known thermodynamic identities $c_p-c_V=T_0\alpha_p^2K_T$ and $c_p/c_V=K_S/K_T$ where $K_S\equiv -V(\partial p/\partial V)_S$ is the adiabatic bulk modulus, the longitudinal heat capacity obeys

\begin{equation}\label{24}
\frac{c_l}{c_V}\,=\,\frac{K_S+4G/3}{K_T+4G/3}\,\,.
\end{equation}
In analogy to the standard abbreviation $\gamma\equiv c_p/c_V$ (a quantity that is also generally frequency dependent) we define

\begin{equation}\label{23}
\gamma_l\,\equiv\,\frac{c_l}{c_V}\,.
\end{equation}
This quantity equals $\gamma$ at low frequencies where $G$ becomes negligible compared to $K_T$, but generally for viscous liquids approaching the glass transition one has $|\gamma_l|<|\gamma|$. The third equation involves the $xx$-component of the stress tensor which by Eq. (\ref{5}) in the one-dimensional case is given by

\begin{equation}\label{25}
\sigma_{xx}\,=\,(K_T+4G/3)\,\pa_x u\,-\,\bv\dt\,.
\end{equation}
The final equation is the defining equation for the time integral $\delta q$ of the heat current $j$:

\begin{equation}\label{26}
\frac{\pa}{\pa t}\delta q\,=\,j\,.
\end{equation}
In terms of $\delta q$ the definition of the heat-conduction coefficient $\lambda$ Eq. (\ref{7}) becomes 
\begin{equation}\label{27}
\delta q\,=\,-\frac{\lambda}{s}\,\pa_x\dt\,.
\end{equation}

In order to solve the four equations Eqs. (\ref{20}), (\ref{21}), (\ref{25}) and (\ref{27}) it is convenient to first rewrite them in dimensionless units. If $l_D$ is the complex number with positive real part defined by (here and henceforth $c_l$ is used in the definition of $l_D$)

\begin{equation}\label{28}
l_D^2\,\equiv\,\frac{\lambda}{c_l\,s}\,,
\end{equation}
the (complex) dimensionless space and time variables are defined by

\begin{eqnarray}\label{29}
\tx\,&\equiv&\,x/l_D\nonumber\\\tit\,&\equiv&\,st\,.
\end{eqnarray}
Any function $f(x,t)$ is made dimensionless by writing $f(x,t)/f_0\equiv\tf(\tx,\tit)$. The four basic fields of the problem are scaled as follows:

\begin{eqnarray}\label{30}
\tu\,&\equiv&\,u/l_D\nonumber\\\dtt\,&\equiv&\,\dt/T_0\nonumber\\\tsig\,&\equiv&\,\sigma_{xx}/K_T\\\dtq\,&\equiv&\,\delta q/(K_Tl_D)\nonumber\,.
\end{eqnarray}
We finally define the following dimensionless linear-response quantities:

\begin{eqnarray}\label{31}
\tc\,&\equiv&\,T_0c_l/K_T\nonumber\\\tg\,&\equiv&\,4G/(3K_T)\\\tal\,&\equiv&\,T_0\alpha_p\nonumber\,.
\end{eqnarray}
For later reference, note that in terms of these variables the following identity applies

\begin{equation}\label{31a}
\frac{\tal^2}{\tc(1+\tg)}\,=\,1-\frac{1}{\gamma_l}\,.
\end{equation}
When Eqs. (\ref{20}), (\ref{21}), (\ref{25}) and (\ref{27}) are rewritten as dimensionless equations, the result is

\begin{eqnarray}
\tu'\,&=&\,\frac{\tal}{1+\tg}\dtt\,+\,a_1\label{34}\\\dtt''\,&=&\,\dtt\,+\,\frac{\tal}{\tc}\,a_1\label{35}\\\tsig\,&=&\,(1+\tg)\,\tu'\,-\,\tal\,\dtt\label{36}\\\dtq\,&=&\,-\tc\,\dtt'\label{37}\,.
\end{eqnarray}
The integration of these equations is straightforward. Substituting Eq. (\ref{34}) into Eq. (\ref{36}) leads to

\begin{equation}\label{38}
\tsig\,=\,(1+\tg)\,a_1\,.
\end{equation}
The integration of Eq. (\ref{35}) introduces two new (frequency-dependent) integration constants:

\begin{equation}\label{39}
\dtt\,=\,e^{\tx}\,a_3\,+\,e^{-\tx}\,a_4\,-\,\frac{\tal}{\tc}\,a_1\,.
\end{equation}
This is substituted into Eq. (\ref{34}), the integration of which results in

\begin{equation}\label{40}
\tu\,=\,\left(1-\frac{\tal^2}{\tc(1+\tg)}\right)\tx\,a_1\,+\,a_2\,+\frac{\tal e^{\tx}}{1+\tg}\,a_3\,-\,\frac{\tal e^{-\tx}}{1+\tg}\,a_4\,.
\end{equation}
Finally, when Eq. (\ref{39}) is substituted into Eq. (\ref{37}) the result is

\begin{equation}\label{41}
\dtq\,=\,-\tc e^{\tx}\,a_3\,+\,\tc e^{-\tx}\,a_4\,.
\end{equation}

The complete solution is summarized as follows

\begin{equation}\label{42}
\begin{pmatrix}\tsig\\ \dtt\\\tu\\\dtq\end{pmatrix}\,=\,\mma(\tx)\begin{pmatrix}a_1\\ a_2 \\ a_3 \\a_4\end{pmatrix}\,,
\end{equation}
where

\begin{equation}\label{43}
\mma(\tx)\,=\,\begin{pmatrix}1+\tg & 0 & 0& 0\\-\frac{\tal}{\tc} & 0 & e^{\tx} & e^{-\tx}\\\left(1-\frac{\tal^2}{\tc(1+\tg)}\right)\tx & 1 & \frac{\tal e^{\tx}}{1+\tg} & -\frac{\tal e^{-\tx}}{1+\tg}\\0 & 0 & -\tc e^{\tx} & \tc e^{-\tx}\end{pmatrix}\,.
\end{equation}
The transfer matrix $\tma(\tx)$ connecting the fields at $\tx=0$ with those at $\tx$,

\begin{equation}\label{44}
\begin{pmatrix}\tsig(\tx)\\ \dtt(\tx)\\\tu(\tx)\\\dtq(\tx)\end{pmatrix}\,=\,\tma(\tx)\begin{pmatrix}\tsig(0)\\ \dtt(0)\\\tu(0)\\\dtq(0)\end{pmatrix}\,,
\end{equation}
is given by Eq. (\ref{13}). The explicit calculation of $\tma(\tx)$ results in 

\begin{equation}\label{45}
\tma(\tx)\,=\,\begin{pmatrix}1 & 0 & 0 & 0\\\frac{\tal(\cosh(\tx)-1)}{\tc(1+\tg)} & \cosh(\tx) & 0 & -\frac{\sinh(\tx)}{\tc}\\\frac{\tal^2(\sinh(\tx)-\tx)+\tc(1+\tg)\tx}{\tc(1+\tg)^2} & \frac{\tal\sinh(\tx)}{1+\tg} & 1 & \frac{\tal(1-\cosh(\tx))}{\tc(1+\tg)}\\-\frac{\tal\sinh(\tx)}{1+\tg} & -\tc\sinh(\tx) & 0 & \cosh(\tx)\end{pmatrix}\,.
\end{equation}
As in the case of simple heat conduction, a direct calculation shows that the consistency requirement $\tma(\tx)\tma(\ty)=\tma(\tx+\ty)$ is fulfilled, implying that $\tma({\rm 0})={\rm 1}$ and $\tma(-\tx)=\tma(\tx)^{\rm -1}$.

\section{The thermal response for a number of boundary conditions}

Consider now a situation where at the surface $x=0$ one imposes a heat current density with amplitude $j(0)$ and measures the corresponding temperature amplitude $\dt(0)$. The liquid is assumed to be fixed at the $x=0$ surface, i.e., $u(0)=0$. The experimentally relevant quantity is the thermal (area-specific) admittance, $Y= j(0)/\dt(0)$. If one defines the dimensionless thermal compliance by

\begin{equation}\label{45b}
\tJ\,\equiv\,\frac{\dtq(0)}{\dtt(0)}\,,
\end{equation}
in real units one has

\begin{equation}\label{45c}
Y\,=\,\frac{K_Tl_Ds}{T_0}\tJ\,.
\end{equation}
We proceed to calculate $Y$ for a number of cases by means of the inverse transfer matrix, 

\begin{equation}\label{45d}
\pma\,\equiv\,\tma^{\rm -1}(\tL)\,=\,\tma(-\tL)\,.
\end{equation}
For reference we give the explicit matrix $\pma$ found by substituting $\tx=-\tL$ into Eq. (\ref{45}):

\begin{equation}\label{45e}
\pma \,=\,
\begin{pmatrix}1 & 0 & 0 & 0\\\frac{\tal(\cosh(\tL)-1)}{\tc(1+\tg)} & \cosh(\tL) & 0 & \frac{\sinh(\tL)}{\tc}\\\frac{\tal^2(-\sinh(\tL)+\tL)-\tc(1+\tg)\tL}{\tc(1+\tg)^2} & -\frac{\tal\sinh(\tL)}{1+\tg} & 1 & \frac{\tal(1-\cosh(\tL))}{\tc(1+\tg)}\\\frac{\tal\sinh(\tL)}{1+\tg} & \tc\sinh(\tL) & 0 & \cosh(\tL)\end{pmatrix}\,.
\end{equation}
For each case we shall explicitly give the low frequency (thermally thin) and high frequency (thermally thick) limits.

\textbf{Case A. Free mechanical termination,} $\tsig(\tL)=0$. 
\\ \textbf{Subcase 1. Adiabatic termination.}
\\ This case corresponds to $\dtq(\tL)=0$ and thus Eq. (\ref{44}) implies

\begin{equation}\label{46}
\begin{pmatrix}\tsig(0)\\ \dtt(0)\\ 0\\\dtq(0)\end{pmatrix}\,=\,\pma\begin{pmatrix} 0\\  \dtt(\tilde L)\\\tu(\tilde L)\\0\end{pmatrix}\,.
\end{equation}
From this we get

\begin{equation}\label{47}
\tJ\, =\,\frac{\dtq(0)}{\dtt(0)}\,=\,\frac{P_{42} \dtt(\tL)+P_{43} \tu(\tL)}{P_{22} \dtt(\tL)+P_{23} \tu(\tL)}\,=\,\frac{P_{42}}{P_{22}}=\tc \tanh(\tL)\,.
\end{equation}
Thus the thermal admittance is given by

\begin{equation}\label{48}
Y\,=\,c_l l_D s \tanh(L/l_D)\,=\,\sqrt{\lambda c_l s}\tanh(L/l_D)\,.
\end{equation}
In the thermally thin limit, $L\ll |l_D|$, this reduces to $Y=\sqrt{\lambda c_l s}L/l_D=c_l L s$. In the thermally thick limit, $L\gg |l_D|$, one finds $Y=\sqrt{\lambda c_l s}$.

\textbf{Subcase 2. Isothermal termination.} 
\\ This case corresponds to $\dtt(\tL)=0$ and thus Eq. (\ref{44}) implies

\begin{equation}\label{49}
\begin{pmatrix}\tsig(0)\\ \dtt(0)\\ 0\\\dtq(0)\end{pmatrix}\,=\,\pma\begin{pmatrix} 0\\  0\\\tu(\tL)\\\dtq(\tL)\end{pmatrix}\,.
\end{equation}
From this we get

\begin{equation}\label{47a}
\tilde J\,=\,\frac{\dtq(0)}{\dtt(0)}\,=\,\frac{P_{43} \tu(\tL)+P_{44} \dtq(\tL)}{P_{23} \tu(\tL)+P_{24} \dtq(\tL)}\,=\,\frac{P_{44}}{P_{24}}\,=\,\tc \coth(\tL)\,.
\end{equation}
Thus the thermal admittance is given by

\begin{equation}\label{48a}
Y=c_l l_D s \coth(L/l_D)=\sqrt{\lambda c_l s}\coth(L/l_D)\,.
\end{equation}

In the thermally thin limit, $L\ll |l_D|$, this reduces to $Y=\sqrt{\lambda c_l s}\,l_D/L=\lambda/L$. In the thermally thick limit, $L\gg |l_D|$, one finds $Y=\sqrt{\lambda c_l s}$.

\textbf{Case B. Clamped mechanical termination,} $\tu(\tilde L)=0$.
\\ \textbf{Subcase 1. Adiabatic termination.}
\\This case corresponds to $\dtq(\tL)=0$ and thus Eq. (\ref{44}) implies

\begin{equation}\label{49a}
\begin{pmatrix}\tsig(0)\\ \dtt(0)\\ 0\\\dtq(0)\end{pmatrix}\,=\,\pma\begin{pmatrix}\tsig(\tilde L)\\ \dtt(\tilde L)\\0\\0\end{pmatrix}\,.
\end{equation}
From the second and fourth equation of Eq. (\ref{49a}) we get

\begin{equation}\label{50}
\tJ\,=\,\frac{\dtq(0)}{\dtt(0)}\,=\,\frac{P_{41} +P_{42} {\dtt(\tL)}/{\tsig(\tL)}}{P_{21} +P_{22} {\dtt(\tL)}/{\tsig(\tL)}}\,,
\end{equation}
and from the third we get $0=P_{31}\tsig(\tL)+P_{32}\dtt(\tL)$ or

\begin{equation}\label{50a}
{\dtt(\tL)}/{\tsig(\tL)}\,=\,-\frac{P_{31}}{P_{32}}\,.
\end{equation}
Substituting this into Eq. (\ref{50}) leads to

\begin{equation}\label{50b}
\tJ\,=\,\frac{P_{41}P_{32}-P_{42}P_{31}}{P_{21}P_{32}-P_{22}P_{31}}\,.
\end{equation}
A calculation of this leads to
\begin{equation}\label{50c}
\tJ\,=\,\frac{\tc}{\coth(\tilde L)+(\gamma_l-1)/\tL}\,.
\end{equation}
Thus in real units \cite{note2}

\begin{equation}\label{51}
Y\,=\,\frac{c_l l_D s}{\coth(L/l_D)+(l_D/ L)(c_l/c_V-1)}\,.
\end{equation}
In the thermally thin limit, $L\ll |l_D|$, this reduces to $Y=c_VLs$. In the thermally thick limit, $L\gg |l_D|$, one finds $Y=\sqrt{\lambda c_l s}$.

\textbf{Subcase 2. Isothermal termination.}
\\This case corresponds to $\dtt(\tL)=0$ and thus Eq. (\ref{44}) implies

\begin{equation}\label{52}
\begin{pmatrix}\tsig(0)\\ \dtt(0)\\ 0\\\dtq(0)\end{pmatrix}\,=\,\pma\begin{pmatrix}\tsig(\tilde L)\\ 0\\0\\\dtq(\tilde L)\end{pmatrix}\,.
\end{equation}
From the second and fourth equation of Eq. (\ref{52}) we get

\begin{equation}\label{53}
\tJ\,=\,\frac{\dtq(0)}{\dtt(0)}\,=\,\frac{P_{41} +P_{44} {\dtq(\tilde L)}/{\tsig(\tilde L)}}{P_{21} +P_{24} {\dtq(\tilde L)}/{\tsig(\tilde L)}}\,,
\end{equation}
and from the third we get 

\begin{equation}\label{54}
{\dtq(\tilde L)}/{\tsig(\tilde L)}\,=\,-\frac{P_{31}}{P_{34}}\,.
\end{equation}
Substituting this into Eq. (\ref{53}) leads to

\begin{equation}\label{55}
\tJ\,=\,\frac{P_{41}P_{34}-P_{44}P_{31}}{P_{21}P_{34}-P_{24}P_{31}}\,,
\end{equation}
or

\begin{equation}\label{55a}
\tJ\,=\,\tc\, \frac{(\gamma_l-1)\sinh(\tL)+\tL \cosh(\tL)}{2(\gamma_l-1)(\cosh(\tL)-1)+\tL \sinh(\tL)}\,.
\end{equation}
Thus in real units

\begin{equation}\label{56}
Y=c_l l_D s\,\frac{(c_l/c_V-1)\sinh(L/l_D)+ (L/l_D) \cosh(L/l_D)}{2(c_l/c_V-1)(\cosh(L/l_D)-1)+ (L/l_D) \sinh(L/l_D)}\,.
\end{equation}
In the thermally thin limit, $L\ll |l_D|$, this reduces to $Y=\lambda/L$. In the thermally thick limit, $L\gg |l_D|$, one finds $Y=\sqrt{\lambda c_l s}$.

In the thermally thin limit ($|l_D(\omega)| \gg L$) the above results are summarized in table \ref{table:Ythin} for the four combinations of thermal and mechanical boundary conditions at the surface $x=L$. It was to be expected that the isothermal boundary condition gives the heat conductivity in both mechanical cases, because heat conductivity does not depend on whether constant volume or constant stress boundary conditions apply. For the adiabatic boundary condition it is also expected that the admittance is related to the constant volume specific heat $c_V$. That the admittance for free mechanical boundary condition is related to $c_l$ and not $c_p$ is not trivial; it relates to the fact that the displacement associated with the temperature oscillation is longitudinal. In the thermally thick limit, $|l_D(\omega)| \ll L$ we se that the thermal admittance becomes $Y=(sc_l \lambda)^{1/2}$ and thus is related to the \textit{longitudinal specific heat for all combinations of boundary conditions considered} (table \ref{table:Ythick}). 

\begin{table}[tb!] 
\begin{tabular}
{|c| c c|}\hline $Y_{\text{thin}}$& adiabatic & isothermal \\ \hline free & $c_lLs$ & $\lambda/L$ \\ clamped & $c_VLs$ & $\lambda/L$ \\ \hline
\end{tabular}
\caption{ The (area-specific) thermal admittance $Y$ at the boundary $x=0$ for different boundary conditions at $x=L$ in the thermally thin limit, $|l_D(\omega)| \gg L$.}\label{table:Ythin}
\end{table}

\begin{table}[tb!] 
\begin{tabular}
{|c| c c|}\hline $Y_{\text{thick}}$& adiabatic & isothermal \\ \hline free &$(sc_l \lambda)^{1/2}$  & $(sc_l \lambda)^{1/2}$ \\ clamped &$(sc_l \lambda)^{1/2}$  &$(sc_l \lambda)^{1/2}$  \\ \hline
\end{tabular}
\caption{ The (area-specific) thermal admittance $Y$ at the boundary $x=0$ for different boundary conditions at $x=L$ in the thermally thick limit ($|l_D(\omega)| \ll L$.}\label{table:Ythick}
\end{table}

\section{Discussion}

How does the result $Y=(sc_l \lambda)^{1/2}$ apply to real effusivity measurements? Are the adiabatic and sliding boundary conditions on faces perpendicular to the heating plane a good approximation? In actual experiments the heating plate width $W$ fulfils $|l_D(\omega)| \ll W$ to assure one-dimensional heat flow orthogonal to the heating plane. The associated displacement field has a range of $|l_D(\omega)|$, and if the solid slab supporting the heating film is much thicker than this length, it will clamp motion along the plane and only allow displacement orthogonal to the plate. In this case the above solution is a good approximation.

One might expect that clamping problems could be circumvented by using an unsupported film free to expand laterally, but that is not correct. Suppose that the film were unsupported and consequently did not stress the liquid. First, we note that there would be a technical problem with the $3\omega$ detection technique since thermal expansion straining the film would give a spurious resistance change on top of the thermally induced resistance change. That problem may be solved, however, and is not the main issue here. Even in this case the thermomechanical coupling, however, would introduce anisotropic stresses: Suppose the strain $\epsilon_{yy}$ were homogeneous on any plane at right angle to the $x$-axis. Then the displacement in the $y$-direction at the boundary of the plate would be proportional to the lateral dimension $W$, but at a distance of the order $|l_D(\omega)|$ in the $x$-direction it would be zero. This would produce enormous shear strains in the liquid (of order $\epsilon_{yy}\frac{W}{|l_D(\omega)|}$), the relaxation of which is controlled by the (high) shear modulus. The condition $|l_D(\omega)| \ll W$ needed for one-dimensional heat flow thus simultaneously gives rise to a situation of partial and anisotropic clamping of the liquid. To determine precisely which effective frequency-dependent specific heat {\it is} actually measured in this (hypothetical) case of a film free to expand laterally requires a detailed calculation; it will not be simply $c_p$.

The ``adiabatic'' method of Christensen's 1985 paper \cite{chr85} is quite different from the Birge-Nagel method, but suffers from the same basic problem that the liquid upon thermally expanding must flow to release the stresses, and that this flow is partly inhibited by the very large viscosity. Thus as for most other calorimeters, the adiabatic method also has uncontrolled mechanical stresses and a stress tensor that is not diagonal. This means that $c_p(\omega)$ is not measured. As mentioned, to determine precisely which quantity a given method measures requires a detailed analysis taking into account the properties of the calorimeter materials. It is quite possible that the discrepancy between the 1985 Birge-Nagel and Christensen measurements on glycerol derives from the two methods {\it de facto} not measuring the same quantity.

Studying the frequency-dependent specific heat via effusion can be done in other geometries than the planar. Birge and Nagel \cite{bir86} also made experiments in an axially symmetric geometry. This was done in order to separate out any possible frequency dependence of the heat conductivity. These experiments were not presented as a function of frequency. Rather, temperature was scanned for fixed frequencies which makes the analysis somewhat indirect (the thermomechanical coupling was not considered in this early experiment, either). Christensen and Olsen \cite{chr98} performed an experiment in spherical symmetry and did take into account the thermomechanical coupling. A small drop of 1,2,4-butanetriol of radius $ r_2=0.35 \text{mm}$ was placed around a temperature-dependent resistor (thermistor) of radius $r_1= 0.15 \text{\,mm}$. By using the $3\omega$-technique the thermal compliance $J_{\rm tot}=Y_{\rm tot}/s$ was found. In the thermally thin limit one has \cite{chr98}

\begin{equation} 
J_{\rm tot}=c_V V\frac{(r_2/r_1)^3 K_S +4/3 G}{(r_2/r_1)^3 K_T +4/3 G}
\end{equation}
where $V$ is the liquid volume. This means that the longitudinal specific heat is measured in the limit $r_2/r_1 \rightarrow 1$ (thin liquid shell) and that the isobaric specific heat is measured in the limit $r_2/r_1 \rightarrow \infty$ (thick liquid shell). These conditions could only be met at very low frequencies, however ( $\leq 40 \text{mHz}$). 

How large is the difference between $c_p$ and $c_l$? From the two identities for $\gamma$ and $\gamma_l$

\begin{equation} 
\gamma\,=\,\frac{c_p}{c_V}=\frac{K_S}{K_T}
\end{equation}
and

\begin{equation} 
\gamma_l\,=\,\frac{c_l}{c_V}\,=\,\frac{M_S}{M_T}
\end{equation}
the following useful result follows

\begin{equation}\label{tagessuperligning}
\frac{c_p-c_l}{c_p}\,=\,\frac{4}{3}\frac{G}{M_T}\frac{c_p-c_V}{c_p}\,.
\end{equation}
For glycerol the relative change in measured specific heat ($45 \%$ at the glass transition) gives a rough upper estimate of $(c_p-c_V)/c_p$. The infinite-frequency shear and bulk moduli are $4$ GPa and $11$ GPa respectively. Thus the factor $4G/3M_T$ is $33 \%$ as an upper estimate. In the case of glycerol the deviation of $c_l$ from $c_p$ becomes $15 \%$ or one third of the relaxation strength. For liquids with larger expansion coefficient, the effect is larger. Note that this is a frequency-dependent correction and that $c_l$ does not just scale with $c_p$ -- it has a different frequency dependence. Equation (\ref{tagessuperligning}) expresses the influence of the thermo-mechanical coupling on $c_l$ in a nutshell: At low frequencies (or high temperatures) $G/M_T$ is small, and $c_l$ becomes $c_p$. At high frequencies (or low temperatures) $(c_p-c_V)/c_p$ is negligible, and $c_l$ again becomes $c_p$. This is why the problem treated in this paper is not encountered for specific-heat measurements on solids or on ordinary liquids. The effect is only significant at the glass transition where upon increasing frequency the first factor [$G/M_T$] increases while the second [$(c_p-c_V)/c_p$] decreases. 

Finally, we wish to emphasize that the effect discussed in this paper has nothing to do with the liquid falling out of equilibrium as the glass transition is approached; the effect is present in the equilibrium liquid well above the calorimetric glass transition temperature. Thus,  even if effusivity measurements are performed in, say, the kHz regime, the effect is present and must be taken into account whenever the alpha relaxation time is of order milliseconds or longer.

\section{Conclusions}

Fundamental physical facts make it difficult to measure the frequency-dependent \textit{isobaric} specific heat for liquids approaching the glass transition because the temperature perturbation of a thermal experiment induces stresses. The high shear modulus slows down the relaxation of the deviatoric part of the stress tensor, giving rise to non-isobaric conditions. In planar geometry this is a problem whether experiments are done in the thermally thin or the thermally thick limits. 

In this paper the one-dimensional thermoviscoelastic problem was solved in terms of a transfer matrix, making it easy find the thermal response for any types of boundary conditions. It was found that in the thermally thick case the \textit{isobaric} specific heat should be replaced by the \textit{longitudinal} specific heat. This result applies to real effusivity measurement to the extent they may be considered one-dimensional. 

In conclusion, wide-frequency specific-heat spectroscopy based on thermal-wave effusion determines the longitudinal specific heat. The isobaric and isochoric specific heat can be calculated subsequently if the frequency-dependent mechanical moduli are known. -- Although we here only studied the thermoviscoelastic problem in the frequency domain, these results also give rise to serious questions regarding the interpretation of time-domain enthalpy-relaxation experiments.

\acknowledgments This work was supported by the Danish National Research Foundation's (DNRF) centre for viscous liquid dynamics ``Glass and Time.''


\begin{thebibliography}{99}
\bibitem{note1} It is a matter of convention whether one uses $\exp(i\omega t)$ or $\exp(-i\omega t)$ for the complex time dependence. The form corresponding to $s=i\omega$ (common, e.g., in electrical engineering literature) is convenient because it saves a lot of minus's. The form corresponding to $s=-i\omega$, on the other hand, is preferable when there are plane waves because when $k>0$ a wave of the form $\exp[i(kx-\omega t)]$ moves in the positive x-direction.
\end{thebibliography}
\end{document}